\documentclass[conference]{IEEEtran}
\IEEEoverridecommandlockouts
\usepackage{cite}
\usepackage{amsmath,amssymb,amsfonts}
\usepackage{algorithmic}
\usepackage{graphicx}
\usepackage{textcomp}
\usepackage{xcolor}
\usepackage{amsmath}
\usepackage{amsfonts}
\usepackage{amssymb}
\usepackage{siunitx}
\usepackage{verbatim}
\usepackage{subfig}
\usepackage{array}
\usepackage{hyperref} 
\usepackage[capitalize]{cleveref}
\usepackage[a4paper,left=1.32cm,right=1.32cm,top=1.91cm,bottom=4.38cm,textwidth=18.44cm]{geometry}
\usepackage[utf8]{inputenc} 
\usepackage[T1]{fontenc}    
\usepackage{hyperref}       
\usepackage{url} 

\def\BibTeX{{\rm B\kern-.05em{\sc i\kern-.025em b}\kern-.08em
    T\kern-.1667em\lower.7ex\hbox{E}\kern-.125emX}}
\begin{document}

\title{Exploring the Dyson Ring: Parameters, Stability and Helical Orbit}

\author{\IEEEauthorblockN{Teerth Raval}
\IEEEauthorblockA{\textit{Mechanical Engineering} \\
\textit{Indian Institute of Technology Hyderabad}
}
\and
\IEEEauthorblockN{Dhruv Srikanth}
\IEEEauthorblockA{\textit{Electrical Engineering} \\
\textit{Indian Institute of Technology Hyderabad}
}
}

\maketitle

\begin{abstract}
A Dyson ring is a hypothetical megastructure, that a
very advanced civilization would build around a star to harness
more of its energy. Satellite propagation is of high priority in
such a vast world where distances could very well be measured
in astronomical units. We analyze the ring’s parameters and
stability and propose a stable helical orbit around the Dyson ring
influenced by the gravity of the Dyson ring and the Sun. Taking
theoretically explainable values for all parameters, we describe
our approach to finding this orbit and present the successful
simulation of a satellite’s flight in this path.


\end{abstract}



\section{Introduction}
The Dyson ring is a hypothetical megastructure, derived from the Dyson Sphere\cite{Dyson}, designed to capture a far larger fraction of a star's energy. It is a ring built around a star with the intent of housing human population on the surface of the ring. The logistics of building such a ring are incomprehensible but theoretically, if such a structure were to exist would the ring be inhabitable?

By maintaining the current Earth-Sun distance as the ring radius, temperature conditions on the surface of the ring would be similar to current temperature on the Earth. Optimal gravity conditions would be achieved by spinning the ring about its axis with a calculated angular velocity such that perceived centrifugal force acts as gravity on the ring's surface. Identifying these conditions as prerequisites for habitability, in this study, we provide a 2-pronged analysis of stability and hence viability of the Dyson ring environment.

Considering the radius and rotation of the ring, first, we explore the parameters and stability of the ring itself. Establishing stability of the ring sets the stage for further exploration in this environment. Majority of human evolution has been made possible due to the existence of satellites orbiting around the Earth. Hence, second, we explore the possibility of finding an orbit in the aforementioned dystopian environment and flying a satellite in orbit around the Dyson ring.

\section{Dyson Ring}

\subsection{Parameters of the Dyson Ring}

A Dyson ring, as explained earlier, is a ring centered around the Sun to harvest a much larger fraction of Solar energy.

\begin{table}[h]
\centering
\begin{tabular}{|c|c|c|}
\hline
\textbf{Quantity} & \textbf{Symbol} & \textbf{Value} \\
\hline
Radius of cross-section of ring & $R_{c}$ & \num{6.371e+6} m \\
Radius of ring & $R$ & $1 \text{ AU} = \num{1.496e+11} m$ \\
Density of ring & $\rho$ & \num{2267} kg/m$^3$ \\
Mass of ring & $m$ & \num{2.725e+29} kg \\
\hline
\end{tabular}
\caption{Assumed values of Dyson Ring parameters}
\label{tab:sample_table}
\end{table}

We take the ring to be a torus, with the cross section of the ring being a circle of radius $R_e$, the radius of the Earth. To live on the ring, its radius must be 1 AU, the present distance between Earth and the sun, giving us temperatures we can survive in. We take the density of the ring to be the density of graphene, as it is an extremely light and strong material, making it a good choice for the ring. Thus the mass of the ring is obtained from
\[
m = \rho V = \rho * (2*\pi*R    )*(\pi*R_{e}^2) 
\]


\subsection{Gravity due to the Dyson Ring}
\label{sec:grav-calc}

Proceeding with the calculation of gravitational field from the Dyson ring, note that we have assumed only the Dyson ring and Sun apply forces on each other and any satellites (no other celestial bodies have an effect).

\noindent Take a small mass $dm$ at an angle of $\theta$ from the x direction.
\[
dm = \lambda R d\theta
\]
where $\lambda$ is mass per unit length of the ring. \\
$\lambda$ can be approximated as $(\rho * 2\pi{R_{e}^2} = \num{2.891e+17})$ as $R_e << R_d$

\noindent The gravitational force $dF$ on a satellite of mass $m$ from this small mass $dm$ is:
\[
\mathbf{dF} = \frac{G \cdot m \cdot dm}{|\mathbf{d^2}|} \times \frac{\mathbf{d}}{|\mathbf{d}|}
\]

\begin{figure}[h]
\centering
\includegraphics[scale=0.7]{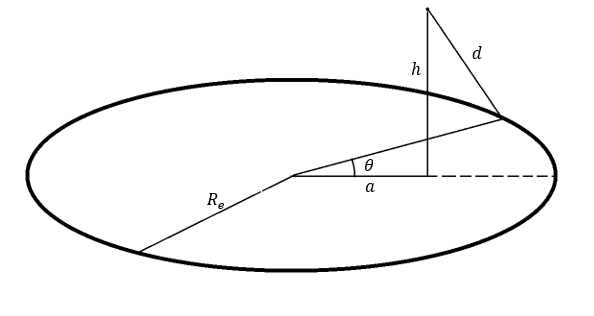}
\caption{\label{fig:grav}Gravitational Force from Dyson ring}
\end{figure}

\noindent Integrating the axial, radial and tangential components of $d\mathbf{F}$ along the entire ring gives us the total gravitational force on the satellite per Cowell's method \cite{Eff_orbit}:
\begin{align*} 
\begin{split}
&F_{rad}(a,h) \\
&= {Gm}\lambda{R_d}\int_{0}^{2\pi} \frac{Rcos\theta - a}{{((Rcos\theta - a)^2 + (Rsin\theta)^2 + (h)^2)}^{3/2}} {d\theta} \\
&F_{tan}(a,h) \\
&= {Gm}\lambda{R_d}\int_{0}^{2\pi} \frac{Rsin\theta}{{((Rcos\theta - a)^2 + (Rsin\theta)^2 + (h)^2)}^{3/2}} {d\theta} \\
&F_{z}(a,h) \\
&= {Gm}\lambda{R_d}\int_{0}^{2\pi} \frac{h}{{((Rcos\theta - a)^2 + (Rsin\theta)^2 + (h)^2)}^{3/2}} {d\theta}
\end{split}
\end{align*}

\noindent On evaluating the integral $F_{tan} = 0$, but the integrals in $F_{rad}$ and $F_{z}$, in the case of a torus/ring, cannot be expressed in the form of elementary functions \cite{Ring_pot}.
Hence we leave them as is and perform numerical integration in code when necessary.

\subsection{Stability of the Ring}

\begin{figure}[h]
    \centering
  \subfloat[\label{stab1}]{%
       \includegraphics[width=0.45\linewidth]{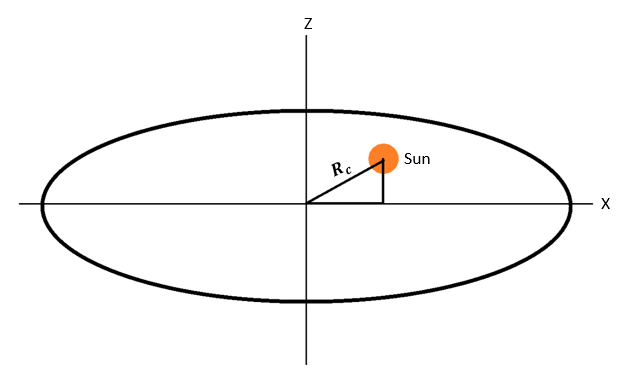}}
    \hfill
  \subfloat[\label{stab2}]{%
        \includegraphics[width=0.45\linewidth]{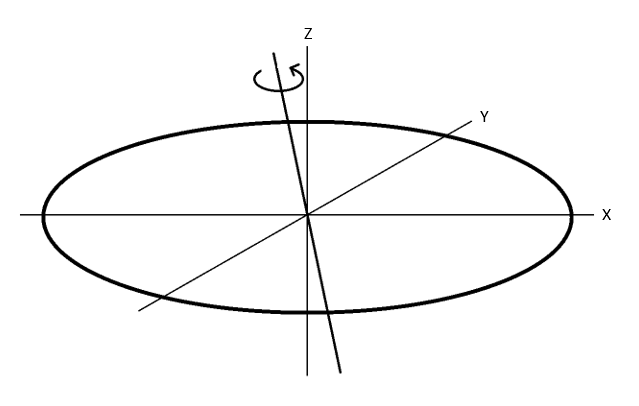}}
    \\
  \caption{(a) Position of Sun relative to ring is off center, (b) Rotation axis deviated from axis of ring.}
  \label{stability} 
\end{figure}

\subsubsection{Deviation of the Ring along Radial or Axial directions}
$\mathbf{F}(0,0) = 0$, hence with the Sun at the centre of the ring it is in equilibrium.

\noindent Force of gravity if deviated along a radial direction is $F_{rad}(a,0)$. Since $\frac{|F_{rad}(a,0)|}{|a|}>0$, if deviated in the plane of the ring, the deviation will keep increasing, leading to the ring and Sun crashing into each other.

\noindent On the other hand if deviated along the axis, the force will be $F_{z}(0,h)$. $\frac{|F_{z}(0,h)|}{|h|}<0$ and thus force will have a restoring effect, leading to oscillation of some sort.

\subsubsection{Deviation of the Ring along any other direction}


\noindent If the Sun is not at the centre of the ring, it could produce a torque due to gravity gradient.

$\vec{R_C}$ is the position vector of the centre of mass (CoM) of the ring relative to the Sun, and $\vec{r}$ is the relative position of an infinitesimal body element, from the CoM of the ring.

\noindent Then moment from the Sun on ring $L_{G}$ is given by:

\begin{align*}
L_{G} &= - \int_{\mathcal{B}}\mathbf{r}\times\frac{GM_e}{{|\mathbf{R}|}^3}{(\mathbf{R_C}+\mathbf{r})}{dm} \\
&= {GM_e}\mathbf{R_C} \times \int_{\mathcal{B}} \frac{\mathbf{r}}{{|\mathbf{R}|}^3} dm 
\end{align*}

Now,
\begin{align*}
{|\mathbf{R}|}^{-3} &= {|\mathbf{R_{C}} + \mathbf{r}|}^{-3} \approx \frac{1}{{r}^3}\left( 1 - \frac{3\mathbf{R_C}\cdot\mathbf{r}}{r^2}\right)
\end{align*}

Thus,
\begin{align*}
L_{G} = {3GM_e}\mathbf{R_C} \times \int_{\mathcal{B}} \frac{1}{r^{5}} {\mathbf{r}}(\mathbf{r}\cdot\mathbf{R_C}) dm \\
\end{align*}

\noindent For small deviations, $\vec{r}$ is approximately constant, equal to radius of the ring $R$. Using this, the property of cross product of vectors $\vec{a} \times (\vec{b} \times \vec{c}) = (\vec{a} \cdot \vec{c}) - (\vec{a} \cdot \vec{b}) \vec{c}$ and the definition of moment of inertia matrix $[I]$ \cite{textbook}, we get

\[
{\mathbf{L}}_G = \frac{{3GM_e}}{R^5}{\mathbf{R}}_C\times[I]{\mathbf{R}}_C
\]

\noindent Inertia matrix $I$ is given by:

\[
I = \begin{pmatrix}
\frac{MR^2}{2} & 0 & 0 \\
0 & \frac{MR^2}{2} & 0 \\
0 & 0 & MR^2
\end{pmatrix}
\]

\noindent Analysing the result for $\mathbf{L}_G$, we see that if the only way that ${\mathbf{R}}_C\times[I]{\mathbf{R}}_C \neq 0$ is if it has a non zero $z$-component and one of the $x$-component or $y$-component be non-zero.

\subsubsection{Precession of the Rotation Axis}
If the axis of rotation is not aligned to axis of the ring, under no external torque, we would observe precession as the ring is axis-symmetric.


Moment of inertia of the ring along the primary axes:

\begin{align*}
I_{x} = I_{y} &= \frac{MR^{2}}{2} ( = I_{T} ) \\
I_{z} &= MR^{2}
\end{align*}

We see that $I_z > I_T$ and thus the ring behaves as an oblate body. Hence retrograde precession is occurs.

\section{God Orbit}

A God orbit is a term we developed for a theoretical multi-purpose orbit that makes full use of the non-uniform environment it is in and where gravitational conditions make the existence of regular Earth-satellite-like orbits difficult.

\begin{figure}[h]
\centering
\includegraphics[scale=0.4]{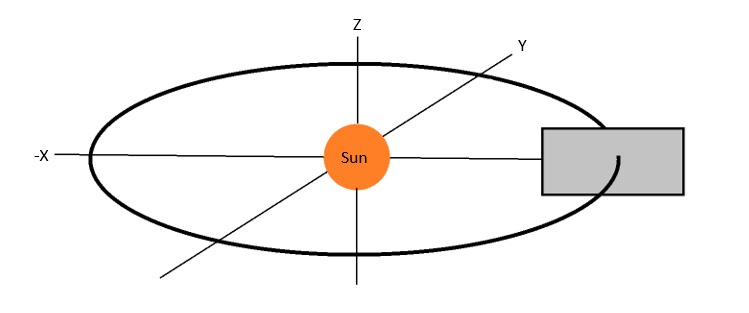}
\caption{\label{fig:cross-sec}Grey X-Z Plane shown here intersects the Ring at (AU, 0, 0).}
\end{figure}

From Section \ref{sec:grav-calc}, we see that the equation for gravitational force is empirically not simplifiable beyond the above-mentioned equation. This complicates the dynamics task as finding orbits is significantly more difficult, let alone simulating motion in orbit. We identified the orbit by visualizing the magnitude of gravitational force to recognize correlations and find possible orbit-conducive patterns. The primary objective is to estimate a path with constant force oriented near tangentially from the path itself that can be used as an orbit.

\subsection{Gravitational Force Visualization}

We identified a possible orbital structure that addressed gravitational force from the Sun as well as gravitational force from the ring:
\begin{enumerate}
    \item Primarily behaving as a satellite of the Sun, the satellite will perform revolutions around the Sun (tangential velocity dependent on Sun's gravity and independent of the ring due to its circular symmetry). We refer to this type of motion from here onwards as \textbf{Revolution.}
    \item Satellite traces a circular path around the cross-section of the ring to sustain enough velocity to evade ring gravity acting as centripetal force. We refer to this type of motion from here onwards as \textbf{Rotation.}
    \item Together, both rotation and revolution move in a helical fashion around the ring. We refer to this type of motion from here onwards as \textbf{Coiling} (since the theoretical orbit forms a coil-shaped structure).
\end{enumerate}

\begin{figure}[h]
\centering
\includegraphics[scale=0.4]{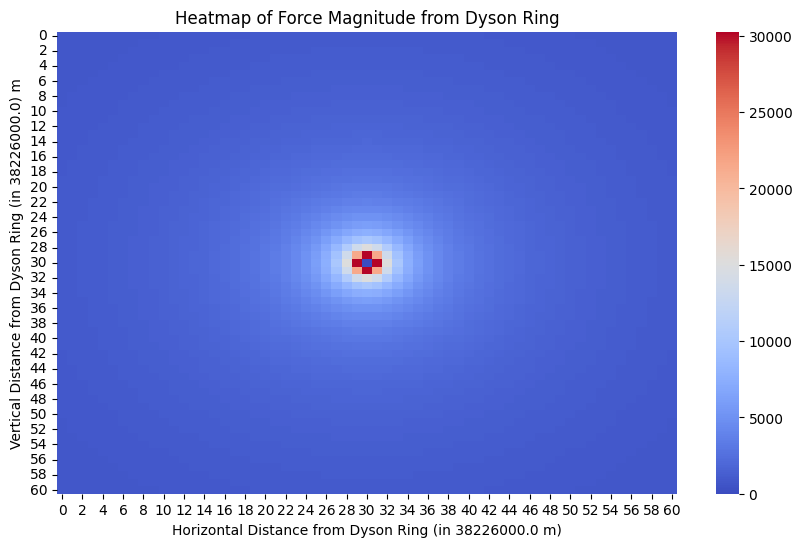}
\caption{\label{fig:heatmap}Gravitational Force Magnitude Heatmap around Dyson Ring in Cross-sectional Plane}
\end{figure}

From detailed exploration of the curve in Desmos\footnote{Desmos file to plot and analyze gravitational force functions: \url{https://www.desmos.com/calculator/sv4q0xepkp}}, we found gravitational force consistencies around the height of medium-earth-orbits from the ring in the cross-sectional plane. We plotted the magnitude of gravitational force in the 2-D Plane (X-Z plane at Y = 0: Cross-section of ring) around the ring over displacement in axial and radial directions as a heatmap where X and Z range from $-6*R_e$ to $6*R_e$. (see Fig. \ref{fig:cross-sec}, the reference for plane representation for Figs. \ref{fig:heatmap} and \ref{fig:rotation}) 

\subsection{Symmetry \& Rotation: Circular Orbit in 2D Plane of Ring's Cross-Section}

From the above heatmap, we find that in this altitude range from the ring, gravitational force is nearly radially symmetrical, allowing us to test circular orbits in this region.

\begin{figure}[h]
\centering
\includegraphics[scale=0.5]{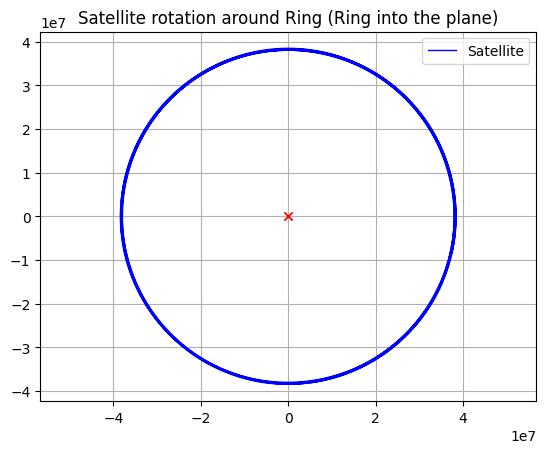}
\caption{\label{fig:rotation} Satellite Orbiting around Ring in Circular path in 2-D Cross-sectional plane (X-Z plane at Y = 0)}
\end{figure}

We experimented with a satellite of mass 1000.0 kg moving in a circular orbit around the ring with an orbit radius of $6*R_e$. gravitational force on the satellite at this range was found to be approximately $1 m/s^{2}$. The initial velocity was found to be approximately $6207.6 m/s$ and if sustained, the satellite would complete a full rotation around the ring in approximately 38691 seconds (10.75 hours)
\footnote{Python code for orbit simulation: \url{https://colab.research.google.com/drive/1AyUygrl_X71k0qGJHLOJWyYuwDUq43M1?usp=sharing}}.

With a minimum time-step of 107 seconds, we simulated the above motion. We obtained that the satellite successfully stays in orbit with minimal deviation from its path even after large number of rotations (see Fig. \ref{fig:rotation}). The centre of the figure is (AU, 0, 0) in Cartesian Sun-centered coordinates where the ring is going into the surface.

[Note: Smaller time-steps provide more accurate satellite path measurements due to finer updates, with which the deviation from the orbit decreased, further showing the orbit's success.]



\begin{figure}[h]
    \centering
  \subfloat[\label{fig:revolution}]{%
       \includegraphics[width=0.45\linewidth]{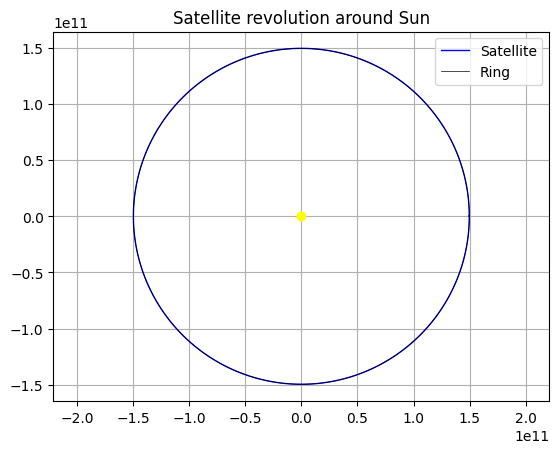}}
    \hfill
  \subfloat[\label{fig:coil}]{%
        \includegraphics[width=0.45\linewidth]{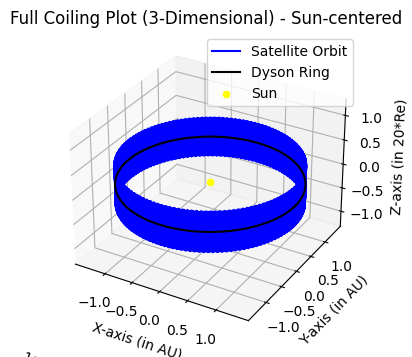}}
    \\
  \caption{(a) Satellite Orbiting around Sun in Circular path similar to Earth's revolution (X-Y plane at Z = 0), (b) 3D Plot of Satellite Coiling around the Dyson Ring}
  \label{orbit-plots} 
\end{figure}

\begin{figure}[h]
\centering
\includegraphics[scale=0.7]{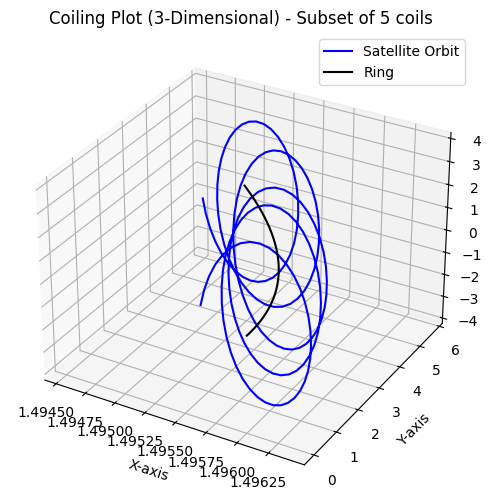}
\caption{\label{fig:coil-zoom} Zoomed-in view of path of Satellite over the first 2.2 days: traces 5 coils}
\end{figure}

We derive the required tangential velocity (to match centrifugal force from gravity) to be approximately $29787 m/s$ with acceleration due to gravity coming out to approximately $0.00593 m/s^2$ radially inward towards the Sun. Traversing an orbit of distance $2*\pi*(1 AU - 6*R_e)$, we estimate the time period for one full revolution to be approximately 31555498.64 seconds (365.22 days). All these parameters are close to the parameters of motion of the Earth around the Sun today. We find that the satellite successfully completes revolutions while staying in orbit (see Fig. \ref{fig:revolution}).

Finally, we simulate both of these motions in tandem: rotation around the ring while revolving around the sun. This gives us a helical pattern - Coiling. In consistent scale, the rotations are invisible to the naked eye. Hence, in the simulation shown in Fig. \ref{fig:coil}, the z-axis is taken in the $1e8$ scale (around $20*R_e$) while x-axis and y-axis are taken in the $1e11$ scale (around $1 AU$). In Fig. \ref{fig:coil-zoom}, we zoom into the path of the satellite over the first 2.2 days ($\textit{192600 seconds} = dt * samples$) and see 5 clearly-defined and well-spaced coils displaying the intricate motion of the satellite.

\section{Simulation Results}

If a Dyson ring with our parameters is built, it would be stable to deviations axially, but unstable in any other direction. Precession would be observed if the axis of rotation of the ring is not aligned to the ring's axis.

Upon theoretical calculation, we determined that geostationary orbits, vertically oscillating Sun-concentric orbits and alternative Earth-satellite-like orbits were impossible due to the system's unique gravitational force and its effect on orbital time period. We used Cowell's method to simulate orbits and found a suitable helical orbit in this new paradigm - one with a circular coil axis (the axis being the same as the Dyson ring). A satellite in this orbit would rotate around the Dyson ring while revolving around the Sun as well. We successfully simulated this coiling of the satellite in the above orbit.

\section{Discussion}

The successful simulation of a coiling orbit around the Dyson ring proves the existence of usable orbits in environments without simple numerically calculable gravitational force. The use of Cowell's method of orbit calculation with the help of heatmaps for force visualization is promising and employable as a preliminary orbit estimation tool in more general scenarios where existing equations are insufficient.

Concerning the above dystopian context, the stability of the spinning Dyson ring (conditional on ring axis alignment) supported by the existence of an employable orbit in this environment bolsters the claim of viability of the Dyson ring as a habitable future home for mankind. Though far into the future, with the fleeting resources of our planet, further exploration into such habitable megastructures could prove to be valuable for the future of mankind. 

We have barely scratched the surface of exploring the possibilities within the field of habitable megastructures and their associated orbits. Fine-tuning the parameter assumptions and experimenting with alternative shapes of the megastructure are valid directions for improvement and future scope. In our calculations, we approximated the torus' exerted gravitational force to that of an infinitely thin ring, due to computational complexity constraints, where its mass per unit length was original calculated based on density of material of the torus. Avoiding the above approximation and defining the torus as such could be an interesting future direction that could help us understand gravitational force equations better.

While Dyson rings themselves are presently hypothetical, our research contributes to the understanding of megastructures that advanced alien civilizations might construct \cite{Other_dyson}. Hence our research could not only help us in the future but also aid in detecting civilizations that might exist and might be advanced enough to create such a megastructure  \cite{station}.

The practical applicability of this study is manifold. Calculation and estimation approaches from this study are directly employable in modern orbit estimation problems, whereas research into habitable megastructures will prove their worth with time. Further exploration into this subject could help us better understand the future of habitable systems and our own race as well as yield interesting insights into fresh approaches for orbit calculation.

\end{document}